\documentclass[a4paper,oneside,12pt]{article}
\usepackage{amsmath}
\usepackage{amsfonts}
\usepackage{amssymb}
\usepackage{amsthm}
\usepackage[dvips]{graphicx}
\usepackage{makeidx}
    \makeindex
\usepackage{layout}

\begin{document}

   \title{Cold atom Clocks and Applications}
    \author{S. Bize,  P. Laurent,  M. Abgrall, \\H. Marion, I. Maksimovic, L. Cacciapuoti, J. Gr\"{u}nert \\ C. Vian, F. Pereira dos Santos, P. Rosenbusch, \\P. Lemonde, G. Santarelli, P. Wolf and A. Clairon
\\ {\it BNM-SYRTE, Observatoire de Paris}\\ {\it 61 Avenue de l'Observatoire 75014 Paris, France.}
\\ A. Luiten, M. Tobar
\\ {\it The University of Western Australia, School of Physics}\\{\it 35
Stirling Highway, Crawley, Western Australia.}
\\ C. Salomon
\\ {\it Laboratoire Kastler Brossel, ENS}\\ {\it 24 rue Lhomond, 75005 Paris,
France}
}
\date{}
\maketitle

\begin{abstract}
This paper describes advances in microwave frequency standards using
laser-cooled atoms at BNM-SYRTE. First, recent improvements of the
$^{133}$Cs and $^{87}$Rb atomic fountains are described. Thanks to
the routine use of a cryogenic sapphire oscillator as an
ultra-stable local frequency reference, a fountain frequency
instability of $1.6\times 10^{-14}\tau^{-1/2}$ where $\tau $ is the
measurement time in seconds is measured. The second advance is a
powerful method to control the frequency shift due to cold
collisions. These two advances lead to a frequency stability of
$2\times 10^{-16}$ at $50~000$~s for the first time for primary
standards. In addition, these clocks realize the SI second with an
accuracy of $7\times 10^{-16}$, one order of magnitude below that of
uncooled devices. In a second part, we describe tests of possible
variations of fundamental constants using $^{87}$Rb and $^{133}$Cs
fountains. Finally we give an update on the cold atom space clock
PHARAO developed in collaboration with CNES. This clock is one of
the main instruments of the ACES/ESA mission which is scheduled to
fly on board the International Space Station in 2008, enabling a new
generation of relativity tests.
\end{abstract}

\section{Introduction: Einstein's legacy in modern clocks}\label{sec:introduction}

Modern clocks using laser cooled atoms owe a great deal to the
famous 1905 "annus mirabilis"  of Einstein. Indeed the three
theoretical problems that Einstein beautifully solved in 1905 are
key ingredients in current atomic clocks, hundred years after
Einstein's work.

\begin{enumerate}
\item First the quanta of light, photons, are routinely used to
cool atoms to microkelvin temperatures and to confine them in
electromagnetic traps. Atom manipulation is a direct application of
energy and momentum exchanges between light and matter. At one
microkelvin, cesium atoms  which form the basis for the current
definition of the SI unit of time, the second, move at an average
speed of 7\,mm.s$^{-1}$, enabling extremely long observation times
and thus precision measurements. On Earth, atomic fountains enable
unperturbed ballistic flight with duration approaching one second.
Furthermore, every experiment in atomic physics routinely uses the
Einstein's photoelectric effect in photodiodes to detect, and
control light beams. Furthermore, the concept of photon is
intimately connected to the famous Planck relationship $E=h \nu$
between energy, Planck's constant and frequency of electromagnetic
radiation, which is of paramount importance in atomic clocks.

\item Second Einstein's theory of brownien motion with the famous
relationship $k_BT=D/\alpha$ between temperature, diffusion
coefficient and friction coefficient not only proved the existence
of atoms, but beautifully applies to Doppler and sub-Doppler laser
cooling mechanisms at work in every cold atom experiment
\cite{Nobel97}. In optical molasses atoms are viscously confined
by the bath of photons, they experience a three-dimensional random
walk in position and storage times in excess of 10 seconds have
been observed for this brownien motion.

 \item Third Einstein's theory of
special (and later general) relativity introduced a new approach
relating space and time, and  the fundamental concept of
relativistic invariance and Lorentz transformation. Einstein
predicted that time in a fast moving frame seems to slow down to
someone not moving with it, and distances appear shorter. This
revolutionary approach had major fundamental as well as practical
consequences in the following century. Clocks in different
reference frames tick at different rates and the well-known GPS
receivers which equip boats, cars, and planes use routinely
Einstein's relativity to determine their position with 10 meter
accuracy. Indeed, the atomic clocks  onboard the 24 GPS satellites
orbiting the Earth at an altitude of 20 000 kms must be corrected
for relativistic effects (time dilation and gravitational shift)
in order to be synchronized with ground clocks and to reach this
positioning accuracy. The correction is about 38 microseconds per
day. If each satellite would not apply this compensation the
positioning error would reach 11 kilometers per day !

These three papers have had revolutionary consequences in Science
and society.

\end{enumerate}

Historically, clocks have played a major role  in tests of
predictions of relativity theories, from the Hafele-Keating clock
transport in jet planes,  the Pound-Rebka gravitational shift
measurement, the Vessot-Levine GP-A Space hydrogen maser
 red-shift measurement, and radar ranging Shapiro delay experiment
\cite{Will93}. In addition the current definition of time in the SI
unit system, the second,  relies on Einstein Equivalence Principle
(EEP). This principle is the foundation for all gravitational metric
theories that describe gravity as a consequence of curved
space-time. The Einstein  Equivalence Principle states
\cite{Will93}:

 \begin{enumerate}
 \item if an uncharged test body is placed at an initial event in
 spacetime and given an initial velocity there, then its
 subsequent trajectory will be independent of its internal
 structure and composition
 \item In any freely falling frame, the outcome of any local non gravitational test
 experiments is independent of the velocity of the frame
 \item the outcome of any local non gravitational test experiment
 is independent of where and when in the universe it is performed
 \end{enumerate}
An immediate consequence of EEP is that the fundamental constants of
physics such as the gravitational constant $G$, or the fine
structure constant $\alpha= e^2/4\pi \epsilon_0 \hbar c$ must be
independent of time and space.

 In this article
we first describe recent progress in the realization of the SI
second using laser cooled cesium and rubidium clocks.  In a second
part we use these highly stable devices to perform new tests of
Einstein Equivalence Principle, namely the constancy of fundamental
constants.

\section{Atomic Fountains}
 The ever increasing control of the motion of atomic samples
is at the origin of recent progress in atomic frequency standards
and precision measurements \cite{Gill2001}. Laser cooled and trapped
atoms enable long observation times required for high precision
measurements. Charged particles confined in Paul or Penning traps
offer extremely long storage enabling high precision mass
measurements, fundamental tests, and the realization of ultra-stable
microwave and optical clocks. The recent NPL frequency measurement
of an optical transition in Sr$^+$ ion with an uncertainty of $3\,
10^{-15}$ \cite{Gill2004} is  only a factor three or four worse than
the current accuracy of cesium fountains. Precision measurements
with neutral atoms on the other hand are usually performed in an
atomic fountain where laser cooled atoms ballistically propagate for
durations up to one second. In the last decade, atomic clocks and
inertial sensors using matter wave interferometry in fountains have
become two of the most important applications of cold atoms
\cite{Gill2001,Clairon1995}. About two dozens of fountain devices
are now used for a variety of applications. It has been shown
recently that microwave and optical clocks as well as matter-wave
inertial sensors belong to the same general class of atom
interferometers \cite{Borde2002}. As an example the current
sensitivity in acceleration measurement with atom interferometers is
on the order of $3\times 10^{-8}\,$m.s$^{-2}$ in one minute
measurement duration and, in a decade, cesium fountain clocks have
gained almost two orders of magnitude in accuracy. As we show in
this paper the fractional inaccuracy of the BNM-SYRTE fountains at
Paris Observatory do not exceed today $7\times 10^{-16}$ which
corresponds to less than a single second error over 50 million
years, allowing for the realization of SI unit of time, the second,
at the same level. About half a dozen fountains throughout the world
at metrology institutes including PTB, NIST, IEN, NPL, have now an
accuracy near $10^{-15}$, making fountains a major contributor to
the accuracy of the TAI (Temps Atomique International). In the
future, many applications, such as positioning systems (GPS,
GALILEO, GLONASS) as well as scientific applications will benefit
from these developments. For instance, deep space satellites have
travel durations of several years across the solar system. Precise
monitoring of their position requires timescales with very low long
term drift. Also, using advanced time and frequency transfer systems
(operating at higher carrier frequency and chip rate and/or using
two way transfer techniques) may lead to positioning accuracy at the
millimeter level for averaging time of a few hundred seconds. This
would impact many geodetic applications.

In this paper we show that prospects for further improvements are
important. A frequency comparison between two fountains exhibits a
stability of $2\, 10^{-16}$ at 50 000 second averaging time, for the
first time for atomic standards. This frequency resolution sets the
stage for clock accuracy at the $10^{-16}$ level for cesium, almost
one order of magnitude potential gain, and even better for rubidium
with its far reduced collision shift \cite{Sortais2000,Fertig2000}.
We begin by recalling the basic operation of fountain atomic clocks
and introduce several new techniques which demonstrate frequency
measurements with a frequency resolution at the $10^{-16}$ level.
The first technique makes use of an ultra-stable cryogenic
oscillator to interrogate the clock transition in the fountain.
Thanks to its extremely good short term frequency stability and low
phase noise, the frequency stability of cesium and rubidium
fountains is one order of magnitude below that of fountains using an
ultra-stable quartz oscillator as interrogation oscillator. It
currently reaches $1.6\times 10^{-14}~\tau^{-1/2}$ where $\tau$ is
the averaging time in seconds. The fundamental quantum noise of the
clock is now reached with atomic samples of up to $10^7$ atoms. The
second advance deals with a new technique to measure and cancel with
high precision the collisional shift in the clock. This shift is a
major plague in cesium clocks and is much reduced (two orders of
magnitude) in rubidium devices \cite{Sortais2000,Fertig2000}. The
method uses interrupted adiabatic population transfer to prepare
precise ratios of atomic densities. We show here that the cesium
collisional shift can be measured and cancelled at the $10^{-3}$
level. By comparing rubidium and cesium fountains over a duration of
six years, a new upper limit for the drift of fundamental constants
has been obtained. Finally we present the development status of the
PHARAO cold atom space clock which is under industrial realization.
PHARAO will fly onboard the international Space station in the frame
of the European ACES mission in 2008-2009 and perform fundamental
physics tests such as an improved measurement of Einstein's
red-shift, search for drift of fundamental constants and special
relativity tests.

\section{Recent advances in cesium and rubidium fountains}\label{sec:developments}
\begin{figure}[htb]
\begin{center}
\includegraphics[height=11cm]{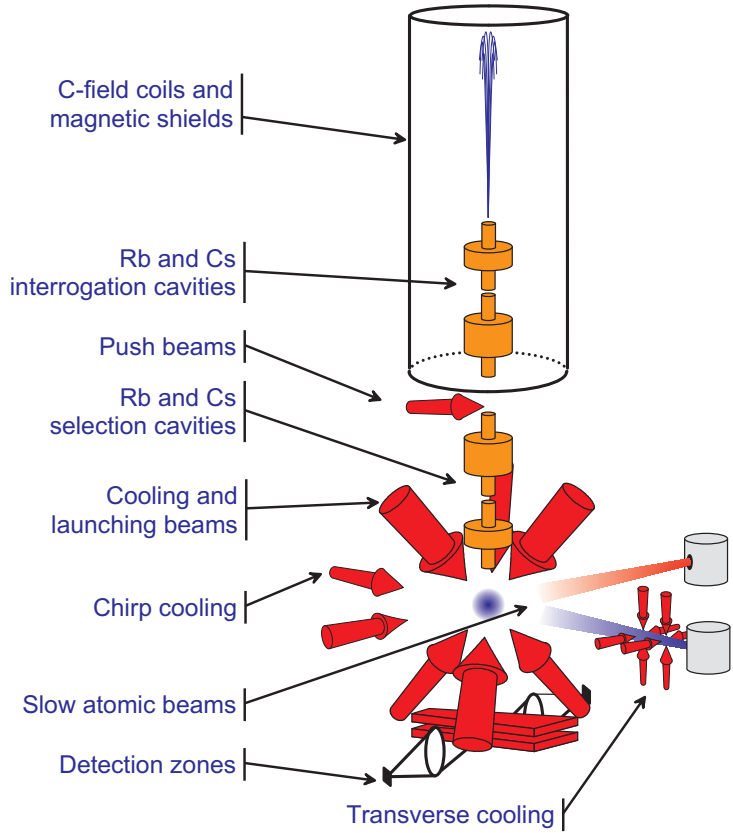}
\end{center}
\caption{Schematic view of the dual Cs-Rb atomic fountain.}
\label{fig:fountain}
\end{figure}

In this section we briefly review recent advances on cesium and
rubidium fountains performed in our laboratory, BNM-SYRTE where
 three laser cooled atomic fountains are in operation. The first one (FO1),
in operation since 1994 \cite{Clairon1995}, has been refurbished
recently. The second one (FOM), a transportable fountain, is derived
from the PHARAO space clock prototype \cite{Laurent1998}. This
fountain was transported on two occasions to the Max Planck
Institute in Garching for direct frequency measurement of the
hydrogen $1s \rightarrow 2s$ transition \cite{Fischer2004}(See
section on stability of fundamental constants). The third one (FO2),
a dual fountain operating with $^{133}$Cs or $^{87}$Rb, is described
in \cite{Sortais2000}. Here we only briefly describe the present
design and recent improvements of FO1 and FO2. A scheme of the
fountain apparatus is shown in Fig.\ref{fig:fountain}. An optical
bench provides through optical fibers all  beams required for
manipulating and detecting the atoms. The fountains operate with lin
$\perp$ lin optical molasses. Atoms are cooled by six laser beams
supplied by preadjusted fiber couplers precisely fixed to the vacuum
tank and aligned along the axes of a 3 dimensional coordinate
system, where the (111) direction is vertical. In FOM, optical
molasses is loaded from a $^{133}$Cs vapor and $3\times 10^7$ atoms
are cooled in 400~ms. In FO1 and FO2, optical molasses are loaded
from a laser slowed atomic beam which is created by diffusing
$^{133}$Cs or $^{87}$Rb vapor through a bundle of capillary tubes.
With this setup $3\times 10^8$ $^{133}$Cs atoms can be loaded in
400~ms in FO1. In FO2 an additional transverse cooling of the atomic
beam increases the loading rate to $10^9$ atoms in 100~ms for
$^{133}$Cs.

The atoms are launched upwards at 4~m.s$^{-1}$ by using moving
optical molasses and cooled to $\sim 1~\mu$K in the moving frame by
adiabatically decreasing the laser intensity and increasing the
laser detuning. In normal operation atoms in the clock level
$|F=3,m_{\mathrm{F}}=0\rangle$ are selected by microwave and light
pulses.

\begin{figure}[htb]
\begin{center}
\includegraphics[height=12cm]{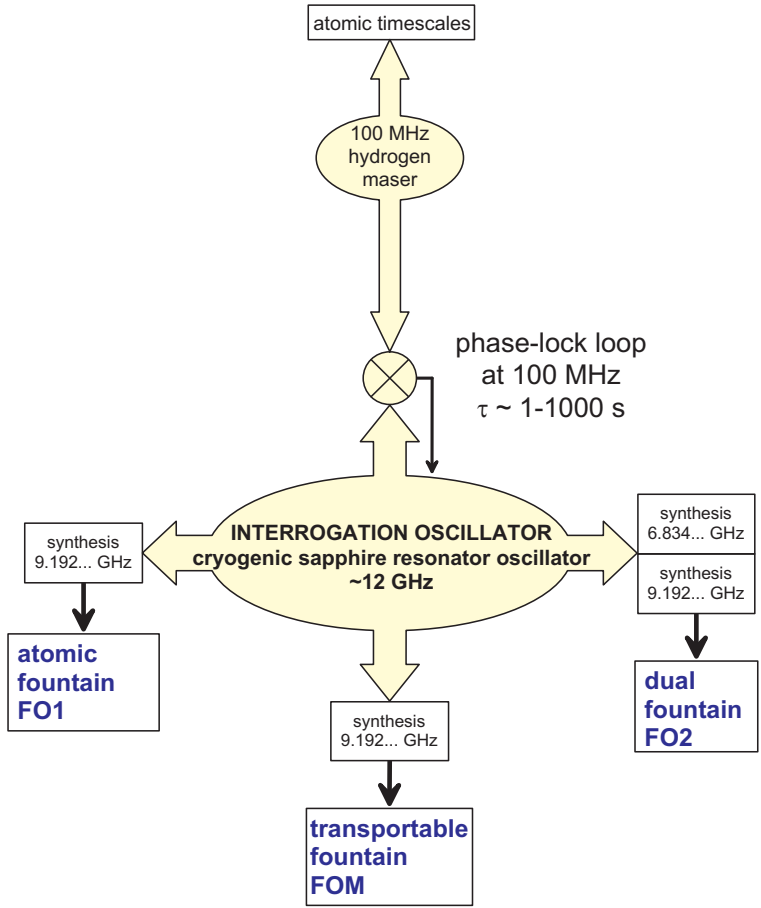}
\end{center}
\caption{BNM-SYRTE fountain ensemble.}
\label{fig:cryogenic_oscillator}
\end{figure}

About 50~cm above the capture zone, a cylindrical copper cavity
(TE$_{011}$ mode)  is  used to probe the hyperfine transition in a
Ramsey interrogation scheme. The cavities have a loaded quality
factor of $Q_{\mathrm{FO1}}=10000$ for FO1 and
$Q_{\mathrm{FO2}}=6600$ for FO2. Both cavities can be fed with two
coupling irises oppositely located on the cavity diameter. Symmetric
or asymmetric feedings are used to evaluate and reduce the residual
Doppler effect due to imperfections of the standing wave in the
cavity and a tilt of the launch direction of the atoms.

\begin{figure}[htb]
\begin{center}
\includegraphics[height=10cm]{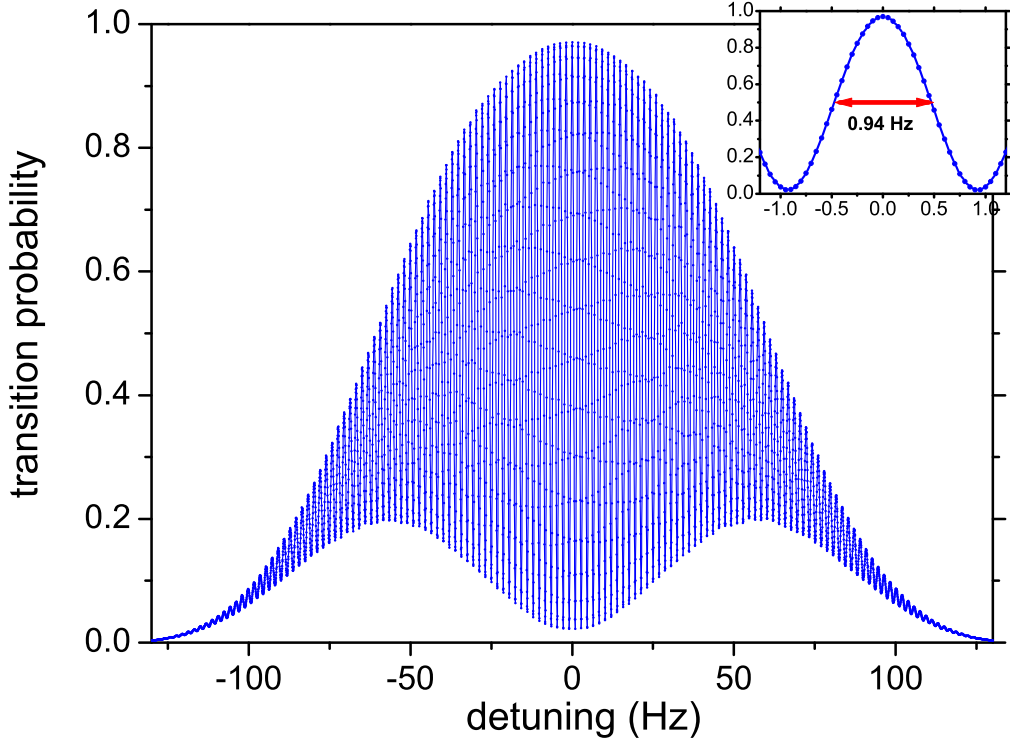}
\end{center}
\caption{Experimental Ramsey fringes (transition probability as a
function of the microwave detuning) measured with $^{133}$Cs in FO2
fountain. The insert shows the central fringe with a FWHM of $\sim
1$~Hz. Each point is a single $1.3$~s measurement.
At half maximum of the central fringe, the signal to noise ratio is
5~000, within $20\%$ of the fundamental quantum noise with $\sim 10^{7}$ detected atoms.} \label{fig:ramsey_Fringes}
\end{figure}

The microwaves feeding the cavities are synthesized from the
signal of an ultra-stable cryogenic sapphire resonator oscillator
(CSO) developed at the University of Western Australia \cite{Mann2001}. As shown
in Fig.\ref{fig:cryogenic_oscillator}, the three fountains use the
same CSO oscillator to synthesize the microwave signals probing the
atomic transition. To reduce its drift, the CSO is weakly
phase-locked to a hydrogen maser. This maser contributes to the
local timescale and TAI (Temps Atomique International) through
various time and frequency transfer systems. With this setup, atomic
fountains are used as primary frequency standard to calibrate TAI
and can be compared to other remote clocks. Nowadays, atomic
fountains are the dominant contributors to the accuracy of TAI.

The 11.932~GHz output signal from the CSO is converted in order to
synthesize 11.98~GHz and 100~MHz signals, both phase coherent with
the H-maser. FO2 uses the 11.98~GHz signal to generate 9.192~GHz
by a home-build low noise synthesizer which achieves a frequency
stability of $3\times 10^{-15}$ at 1~s  by operating only  in the
microwave domain. This scheme reduces at the minimum the phase
noise and the spurious side-bands induced by the down conversion
process. A similar setup is used to synthesize the 6.834~GHz
required for the FO2 fountain operation with $^{87}$Rb. The 150~m
distance between FO1, FOM and the CSO prevents the direct use of
the 11.98~GHz signal. Instead, a 100~MHz signal is synthesized
from the CSO and distributed to FO1 and FOM via a high stability
RF cable. Finally, a 100~MHz to 9.192~GHz  home-made synthesizer
generates the interrogation signal. These additional steps degrade
the phase noise of the interrogation signal in FO1 and FOM with a
frequency stability currently limited to  $\sim 2\times10^{-14}$
at 1~s.

\subsection*{Frequency stability}

\begin{figure}[htb]
\begin{center}
\includegraphics[height=8cm]{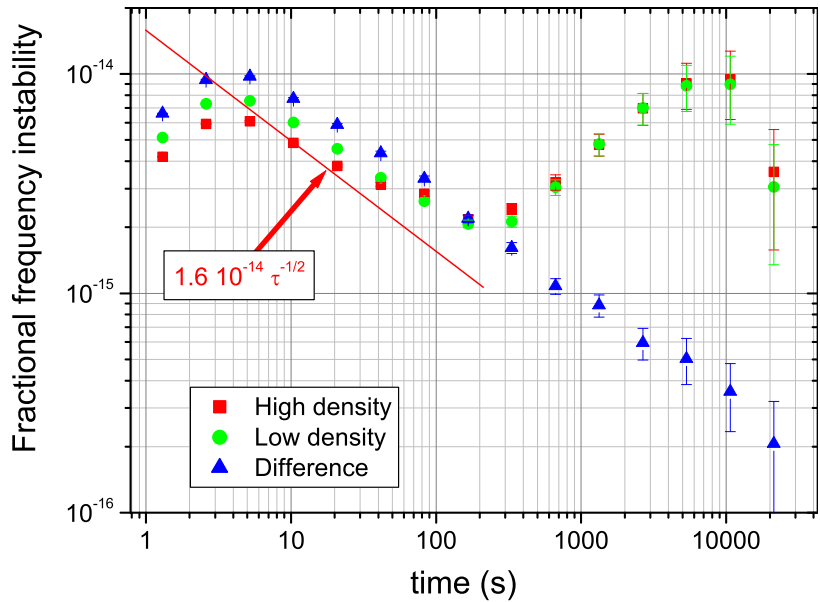}
\end{center}
\caption{Fractional frequency instability of FO2 against CSO for
high density (HD, red squares) and low density (LD, green circles)
configurations. It demonstrates a stability of $1.6\times
10^{-14}\tau^{-1/2}$ for a $^{133}$Cs primary standard. Also shown
is the fractional frequency instability for the differential
measurement between HD and LD (blue triangles). This curve
demonstrates an excellent rejection of the CSO fluctuations in the
differential measurement, allowing for a fractional frequency
resolution of $2.5\times 10^{-16}$ at 20~000~s. In this measurement,
the collisional shift at LD is the frequency difference between HD
and LD ($\sim 5\times 10^{-14}$). It is obtained with a resolution
close to 2 parts in $10^{-16}$ and it is stable at the $0.5\%$ level
over 20~000~s.} \label{fig:FO2HDBD}
\end{figure}
Atoms selected in $|F=3,m_F=0\rangle$ cross the microwave cavity on the
way up and on the way down, completing the two Ramsey interactions.
After the Ramsey interrogation, the populations $N_e$
and $N_g$ of both clock levels $|e\rangle$ and $|g\rangle$ are
measured by fluorescence. The number of detected atoms is typically
$0.5\%$ of the initially captured atoms. The signal $p=N_e/(N_e+N_g)$ is
equal to the atomic transition probability and is insensitive to
atom number fluctuations. A typical Ramsey resonance is presented in
Fig.\ref{fig:ramsey_Fringes}. From the transition probability,
measured on both sides of the central Ramsey fringe, we compute an
error signal to lock the microwave interrogation frequency to the
atomic transition using  a digital servo loop. At the quantum limit
one expects $S/N= 1/\sigma_{\delta p}= 2\sqrt{N}$ for $N$ detected
atoms, where $\sigma_{\delta p}$ is the shot to shot standard
deviation of the fluctuations of the transition probability.
The frequency corrections are applied to a computer controlled high
resolution DDS synthesizer in the microwave generator. These
corrections are used for the accuracy and frequency stability
evaluations of each fountain. The fractional frequency instability
of the FO2 fountain operating with $\sim 10^7$  detected atoms and
measured against the cryogenic oscillator is plotted
Fig.\ref{fig:FO2vsFO1}. At the quantum limit one expects a frequency
instability, characterized by the fractional Allan standard
deviation, given by: $\sigma_{\mathrm{y}}(\tau)= (1/\pi
Q_{at})\sqrt{T_{\mathrm{c}}/N\tau}$, where $Q_{at}\sim 10^{10}$ is
the atomic quality factor, $\tau$ and $T_{\mathrm{c}}$ are
respectively the averaging time and the cycle duration. Above the
servo-loop time constant ($\sim 3$~s) and below 100~s, the
fractional instabilities of FO1 and FO2 are
$\sigma_{\mathrm{y}}(\tau)=2.9\times 10^{-14}\tau^{-1/2}$ and
$1.6\times 10^{-14}\tau^{-1/2}$ respectively, within $\sim 20\%$ of
the standard quantum limit. For longer averaging time the frequency
instability is dominated by the frequency fluctuations of the CSO
and the H-maser. This is the first demonstration of routinely
operated primary frequency standards with frequency instabilities in
the low $10^{-14}\tau^{-1/2}$ region. We will show below that this
excellent short term stability enables an evaluation of systematic
frequency shifts and frequency comparisons between clocks at the
$10^{-16}$ level in a few days.

\subsection*{Accuracy}
All  known systematic frequency shifts are evaluated in our
fountains.
The accuracy budget for each shift is given in table I for
$^{133}$Cs. The overall uncertainty,  the quadratic sum  of all
uncertainties is $7.5\times 10^{-16}$ for FO1,  $6.5\times
10^{-16}$ for FO2 and $8\times 10^{-16}$ for FOM. In the
following, we only discuss some of the most bothersome effects and
the recent improvements in their evaluation. A more complete
discussion of systematic effects can be found in \cite{Vian2004}.

\begin{table}
\renewcommand{\arraystretch}{1.3}
\caption{Systematic fractional frequency shifts for FO1 and FO2.}
\label{tb:accuracy}
\begin{center}
\small
\begin{tabular}{|l|r|r|r|}
\cline{2-4}
\multicolumn{1}{l|}{}&{\small  FO1~($\times 10^{16}$)}& {\small FO2~($\times 10^{16}$)}& {\small FOM~($\times 10^{16}$)}\\
\cline{2-4} \hline
{\small Quadratic Zeeman effect}& $1199.7\pm4.5$ & $1927.3\pm0.3$ & $351.9\pm 2.4$\\
\hline
{\small Blackbody radiation}& $-162.8\pm2.5$ & $-168.2\pm2.5$& $-191.0\pm2.5$\\
\hline
{\small Collisions and cavity pulling (HD)}& $-197.9\pm2.4$ & $-357.5\pm2.0$ & $-34.0\pm5.8$\\
\hline
{\small Spectral purity \& leakage}& $0.0\pm3.3$ & $0.0\pm4.3$& $0.0\pm2.4$\\
\hline
{\small First order Doppler effect}&$<3$&$<3$&$<2$\\
\hline
{\small Ramsey \& Rabi pulling}&$<1$&$<1$&$<1$\\
\hline
{\small Microwave recoil}&$<1.4$&$<1.4$&$<1.4$\\
\hline
{\small Second order Doppler effect}& $<0.08$&$<0.08$&$<0.08$\\
\hline
{\small Background collisions}& $<1$& $<1$& $<1$\\

 \hline \hline
{\small Total uncertainty} & $\pm7.5$ & $\pm6.5$& $\pm7.7$\\
\hline
\end{tabular}
\end{center}
\end{table}

\subsubsection*{Cold collisions and cavity pulling}
The cold collision frequency shift is known to be particularly large
for $^{133}$Cs \cite{Gibble1993,Ghezali1996}. For instance, when FO2
is operated at its best frequency stability the shift is $\sim
10^{-13}$. The linear extrapolation of this effect to zero density
is known to be delicate. As pointed out in the first paper observing
the cold collision shift in $^{133}$Cs fountains \cite{Gibble1993},
selecting atoms in the clock levels using microwaves may lead to
distortions of the position or velocity distribution. Methods to
cope with these effects have been proposed \cite{Fertig2000}, yet
the linear extrapolations have proved to be valid only at the 5\% to
10\% level.

To evaluate the collision shift at the $10^{-3}$ level (a
requirement for a frequency stability and accuracy at $10^{-16}$),
we recently developed a new method based on interrupted adiabatic
passage to select atoms in the $|F=3,m_{\mathrm{F}}=0\rangle$ state
\cite{Pereira2002}. Atomic samples are prepared by transferring
atoms from $|F=4,m_{\mathrm{F}}=0\rangle$ to
$|F=3,m_{\mathrm{F}}=0\rangle$ with an efficiency precisely equal to
$100\%$ (high density, HD) or $50\%$ (low density, LD). With this
method, the atom number is changed without affecting either the
velocity or the position distributions. Therefore, the density ratio
LD/HD is equal to the atom number ratio and is $1/2$ at the
$10^{-3}$ level. Since the collisional shift is proportional to the
atomic density, it can be extrapolated to zero density with this
accuracy. In addition, with this method, the cavity frequency
pulling \cite{Sortais2000,Fertig2000,Bize2001} is also accounted
for.

\begin{figure}[htb]
\begin{center}
\includegraphics[height=7cm]{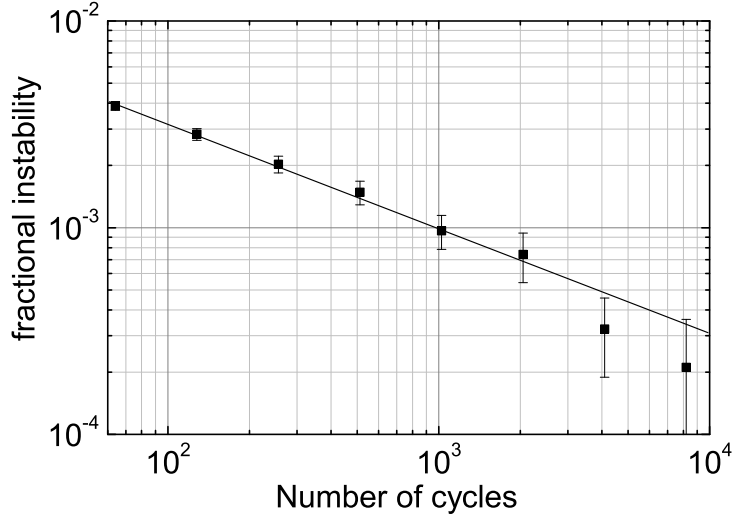}
\end{center}
\caption{Fractional instability of the ratio of the detected atom
number in $|F=4,m_F=0\rangle$ between low density and high density
configurations as a function of the number of fountain cycles. The
measured ratio is $0.5005(2)$. Each cycle lasts $\sim 1.3$~s. The
stability (solid line) decreases as the square root of the number of
cycles.} \label{fig:ratioHDLDFO1}
\end{figure}

The collisional shift is measured in real-time with the following
differential method. The clock is operated alternately in the HD
configuration for $60$~s and in the LD configuration for the next
$60$~s. This timing choice minimizes the noise due to frequency
instabilities of the CSO oscillator. As seen in
Fig.\ref{fig:FO2HDBD}, at $120$~s the stability of FO2 against CSO
is near its minimum. Also, over this $120$~s period, density
fluctuations do not exceed $\sim 1\%$. One the other hand, due to
slow changes in the clock environment, we observe that the density
may fluctuate up to $10-20\%$ over one or several days. Our
differential method efficiently cancels these slow daily density
variations.

In \cite{Pereira2002}, our calculations predicted that the
interrupted adiabatic passage method does provide a LD/HD ratio
precisely equal to 1/2 to better than $10^{-3}$. Initially, we were
experimentally able to realize this ratio at the $1\%$ level.
Improvements in the accuracy of the microwave frequency synthesis
for the adiabatic passage enable us to now reach a precision of
$2\times 10^{-3}$ for this ratio.

During routine operation of the fountains, the number of detected
atoms in each hyperfine state is recorded for both LD and HD
configurations. As seen in Fig.\ref{fig:ratioHDLDFO1}, the Allan
standard deviation of the measured LD/HD atom number ratio decreases
as the square root of the number of fountain cycles (or time), down
to a few parts in $10^4$ for one day of averaging. Despite the
$10-20\%$ slow drift in atom number over days, this ratio remains
remarkably constant. The LD/HD atom number ratio in
$|F=4,m_{\mathrm{F}}=0\rangle$ is found equal to $1/2$ to better
than $10^{-3}$. This method relies on fluorescence measurements made
in the detection zones for each fountain cycle. Various measurements
have been performed to establish that the measurement of this ratio
is not biased by more than $10^{-3}$ due optical thickness effects
in the detection.
 On the other hand,
the LD/HD atom number ratio in $|F=3,m_{\mathrm{F}}=0\rangle$ is
found to slightly differ from 1/2 by $0.3\%$ typically. This
deviation originates from atoms in the $|F=3,m_{\mathrm{F}}\neq
0\rangle$ states populated by imperfections in the state
preparation. This deviation must be taken into account in the
evaluation of the collisional shift. In \cite{Marion2004}, we showed
that the frequency shift of the clock transition due to
$|F=3,m_{\mathrm{F}}\neq 0\rangle$ atoms is at most $1/3$ of that of
collisions between $|F=3,m_{\mathrm{F}}=0\rangle$ and
$|F=4,m_{\mathrm{F}}=0\rangle$ clock states. Their contribution to
the collisional frequency shift is thus at the $0.1\%$ level. In
summary, when this adiabatic passage method is used, we take a
$2\times 10^{-3}$ relative uncertainty for the determination of the
high density cold collision shift.

\subsubsection*{Effect of microwave spectral purity and leakage}

Spectral impurities of the interrogation signal and microwave
leakage may cause shifts of the clock frequency. In order to
evaluate these effects, we make use of their dependance with the
microwave power. We alternate the Ramsey interrogation between a
configuration of $\pi/2$ and $3\pi/2$ pulses, i.e. a variation of a
factor of 9 in microwave power. Within the resolution of the measurement of
$3.3\times 10^{-16}$, no frequency shift is observed. In this
measurement, four data sets are recorded, LD and HD at $\pi/2$ and
LD and HD at $3\pi/2$. In this way, the collisional shift (which may also change with the microwave power) is
evaluated and cancelled for both $\pi/2$ and $3\pi/2$ configurations
by the differential method described above, allowing for the
extraction of a possible influence of microwave spectral purity and
leakage alone.

\subsubsection*{Residual first order Doppler effect}
A frequency shift due to the first order Doppler effect can occur if
the microwave field inside the interrogation cavity exhibits a phase
gradient and the atoms pass the cavity with a slight inclination
from the cavity axis. We determine the frequency shift due to the linear component of the phase gradient in a
differential measurement by coupling the microwave interrogation
signal ``from the left'', ``from the right'' or
symmetrically into the cavity, providing 3 data sets. The observed
shift between the ``left'' and symmetric configuration is
$(-25.3\pm 1.1)\times 10^{-16}$ while the shift between the
``right'' and symmetric configuration is $(+24.0\pm 1.2)\times
10^{-16}$. The magnitude of this residual first order Doppler effect
is consistent with a simple estimate of the residual traveling wave
component in the cavity \cite{Schroder2004} together with a
misalignment between the local gravity and the launch direction
$\lesssim 1$~mrad. The mean of these two measurements is $(-0.7\pm
0.8)\times 10^{-16}$ and consistent with zero, indicating that the
traveling wave component is well cancelled in the symmetric coupling
configuration. Using the atoms as a probe, we can indeed ensure
that the cavity is fed symmetrically to better than $1\%$ in amplitude
and $60$~mrad in phase, which cancels the effect of linear phase gradient
to $\sim 1\%$, better than the above measurement resolution.
As a consequence, only the quadratic phase dependence
of the microwave field remains as a possible source of residual Doppler shift.
A worst case estimate based
on \cite{Schroder2004} gives an upper bound for the fractional
frequency shift of 3 parts in $10^{16}$, which we conservatively
take as the overall uncertainty associated with residual first order
Doppler effect.

Other contributions to the accuracy budget are listed in Table
\ref{tb:accuracy}. The total accuracy currently reaches $7.5$ parts
in $10^{16}$ for FO1, $6.5$ parts in $10^{16}$ for FO2, and $8.0$
parts in $10^{16}$ for FOM. This represents a one order of magnitude
improvement over uncooled cesium devices. In the future, we
anticipate that the extensive use of the methods described above
will enable us to bring the accuracy of $^{133}$Cs fountains below 2
parts in $10^{16}$ and the accuracy of $^{87}$Rb to a even lower
value thanks to its 100-fold lower collision shift
\cite{Sortais2000,Fertig2000}.

\subsubsection*{Frequency comparisons between two $^{133}$Cs fountains at  $2\,10^{-16}$}

\begin{figure}[htb]
\begin{center}
\includegraphics[height=9cm]{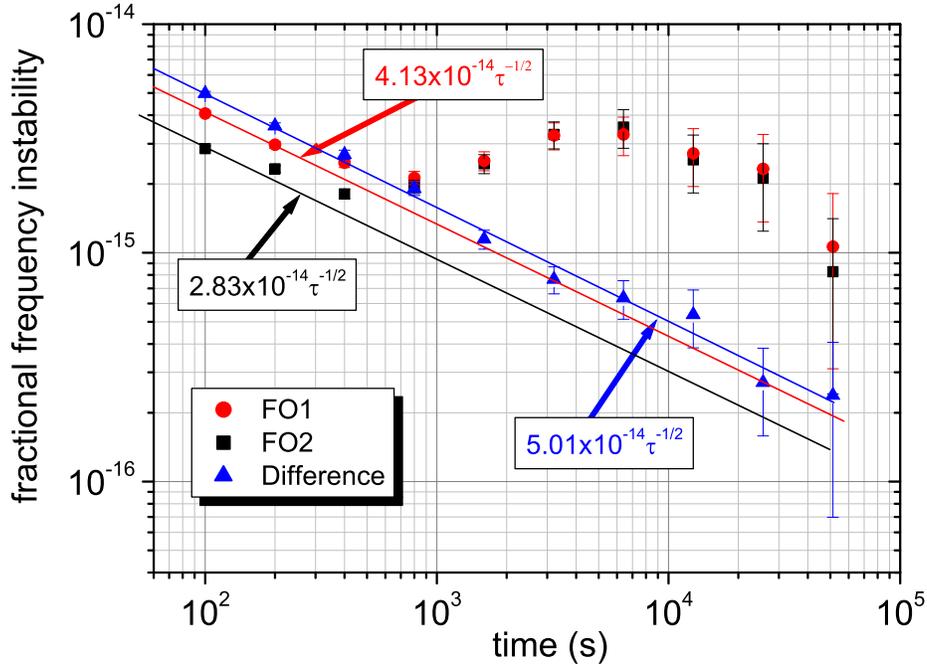}
\end{center}
\caption{Fractional frequency instability (Allan deviation) between FO1 and FO2
fountains (blue triangles). After $50~000$~s of averaging, the stability
between the two fountains is  $2.2$ parts in $10^{16}$. Also
plotted is the fractional frequency instability of FO1 (red circles)
and FO2 (black squares) against the CSO locked to the hydrogen maser.}
\label{fig:FO2vsFO1}
\end{figure}

The routine operation of two atomic fountains near the quantum noise
limit using the CSO as an interrogation oscillator enables frequency
comparisons in the low $10^{-16}$ range, for the first time for
primary frequency standards. Figure \ref{fig:FO2vsFO1} presents the
frequency stability between FO1, FO2 and CSO. Each fountain is
operated in differential mode in order to permanently evaluate and
cancel the collision shift. Appropriate post-processing of the data
thus enables us to construct for each fountain, a clock which is
free of the cold collision shift and whose stability is shown in
Fig. \ref{fig:FO2vsFO1} against the CSO oscillator. Fig.
\ref{fig:FO2vsFO1} also shows that the  combined stability between
these two clocks reaches $2.2\times 10^{-16}$ at $50~000$~s, a
previously unattained long term stability. From this data, we infer
that at least one of the two fountains has a stability below $
(2.2/\sqrt{2})\times 10^{-16}=1.6 \times 10^{-16}$ at the same
averaging time. The mean fractional frequency difference between the
two fountains is $4\times 10^{-16}$, fully compatible with the
accuracy of each of the two clocks as stated in Table
\ref{tb:accuracy}. This very good stability sets a new challenge for
time and frequency transfer systems between remote clocks. As an
example, long distance frequency comparisons between PTB and NIST
fountains were performed at the level of only $6\times 10^{-16}$
after 2 weeks of averaging with GPS \cite{Parker2001}. Similarly,
comparisons between BNM-SYRTE and PTB recently achieved $2\times
10^{-15}$ for one day of integration with TWSTFT \cite{Richard2004}.

\section{Einstein Equivalence Principle and Stability of fundamental constants}\label{sec:alpha}

Highly accurate atomic clocks offer the possibility to perform
laboratory tests of possible variations of fundamental constants.
Such tests interestingly complement experimental tests of the
Local Lorentz Invariance and of the Universality of free-fall to
experimentally establish the validity of Einstein's Equivalence
Principle (EEP). They also complement tests of the variability of
fundamental constants on different timescales, geological
timescale \cite{Damour1996,Fujii2003} and cosmological timescale
\cite{Webb2001,Srianand2004}. Nearly all unification theories (in
particular string theories) violate EEP at some level
\cite{Marciano1984,Damour1994,Damou2002} which strongly motivates
experimental searches for such violations.

    Tests described here are based on highly
    accurate comparisons of atomic energies.
    In principle, it is possible to express any atomic energy as a
    function of the elementary particle properties and the
    coupling constants of fundamental interactions using Quantum
    Electro-Dynamics (QED) and Quantum Chromo-Dynamics (QCD). As a
    consequence, it is possible to deduce a constraint to the
    variation of fundamental constants from a measurement of the stability of the ratio between
    various atomic frequencies.

    Different types of atomic transitions are linked
    to different fundamental constants. The hyperfine
    frequency
    in a given electronic state of alkali-like atoms (involved for instance in $^{133}$Cs, $^{87}$Rb \cite{Marion2003},
    $^{199}$Hg$^{+}$ \cite{Prestage1995,Berkeland1998}, $^{171}$Yb$^{+}$ microwave clocks) can be approximated by:
    \begin{equation}\label{eq:hyperfine_energy}
    \nu_{\mathrm{hfs}}^{(i)} \simeq R_\infty
    c \times\mathcal{A}_{\mathrm{hfs}}^{(i)} \times
    g^{(i)}\left(\frac{m_e}{m_p}\right)~\alpha^2~F_{\mathrm{hfs}}^{(i)}(\alpha),
    \end{equation}
    where the superscript $(i)$ indicates that the quantity depends
    on each particular atom. $R_\infty$ is the Rydberg constant, $c$ the speed of
    light, $g^{(i)}$ the nuclear g-factor, $m_e/m_p$ the electron to
    proton mass ratio and $\alpha$ the fine structure constant.
    In this equation, the dimension is given by $R_\infty
    c$, the atomic unit of frequency. $\mathcal{A}_{\mathrm{hfs}}^{(i)}$ is a numerical factor which depends on
    each particular atom. $F_{\mathrm{hfs}}^{(i)}(\alpha)$ is a relativistic correction
    factor to the motion of the valence electron in the vicinity of the nucleus. This factor
    strongly depends on the atomic number $Z$ and has a major
    contribution for heavy nuclei. Similarly, the frequency of an
    electronic transition (involved in H \cite{Niering2000}, $^{40}$Ca \cite{Helmcke2003},
    $^{199}$Hg$^{+}$ \cite{Bize2003}, $^{171}$Yb$^{+}$ \cite{Stenger2001,Peik2003} optical clocks) can be approximated by
    \begin{equation}\label{eq:electronic_energy}
    \nu_{\mathrm{elec}}^{(i)} \simeq R_\infty
    c \times \mathcal{A}_{\mathrm{elec}}^{(i)} \times F_{\mathrm{elec}}^{(i)}(\alpha).
    \end{equation}
    Again, the dimension is given by $R_\infty c$.
    $\mathcal{A}_{\mathrm{elec}}^{(i)}$ is a numerical factor.
    $F_{\mathrm{elec}}^{(i)}(\alpha)$ is a function of $\alpha$ which
    accounts for relativistic effects, spin-orbit couplings and many-body effects \footnote{It should be noted that in general the
    energy of an electronic transition has in fact a
    contribution from the hyperfine interaction. However, this
    contribution is a small fraction of the total transition energy and thus carries no significant sensitivity to a variation of fundamental constants.
    The same applies to higher order terms in the expression of the hyperfine energy (\ref{eq:hyperfine_energy}). A precision of 1 to 10 $\%$
    on the sensitivity is sufficient to interpret current experiments.}.

    According to \cite{Flambaum2003,Flambaum2004}, the sensitivity to g-factors
    $g^{(i)}$ and to the proton mass $m_p$ can be related to a sensitivity to fundamental parameters,
    namely the mass scale of QCD $\Lambda_{\mathrm{QCD}}$ and the quark
    masses $m_{q}=(m_{u}+m_{d})/2$ and $m_s$. Therefore, any
    measurement of the ratio between atomic frequencies can be interpreted as testing the
    stability of 4 dimensionless fundamental constants: $\alpha$,
    $m_{q}/\Lambda_{\mathrm{QCD}}$, $m_{s}/\Lambda_{\mathrm{QCD}}$ and
    $m_{e}/\Lambda_{\mathrm{QCD}}$. The sensitivity to
    $m_{s}/\Lambda_{\mathrm{QCD}}$ is relatively weak compared to
    the 3 other constants. The sensitivity coefficients have now been calculated for a large
    number of atomic species used in atomic clocks
    \cite{Prestage1995,Flambaum2003,Dzuba1999a,Dzuba1999,Dzuba2000,Karshenboim2000b,Dzuba2003,Flambaum2004,Angstmann2004}.
    Reliable knowledge of these sensitivity coefficient at the
    1$\%$ to $10\%$ level is required to deduce limits to a
    possible variation of each of these fundamental parameters by
    combining the results of several complementary clock comparisons.

\begin{figure}[htb]
\begin{center}
\includegraphics[height=8cm]{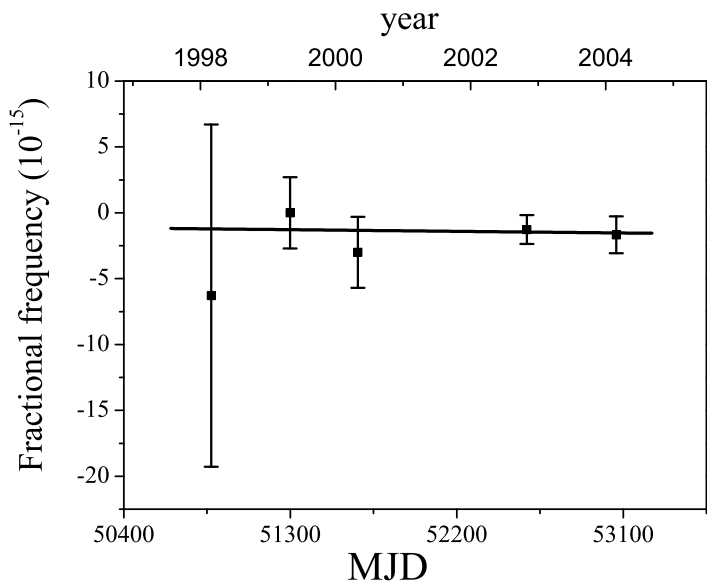}
\end{center}
\caption{Measured $^{87}$Rb frequencies referenced to the $^{133}$Cs
fountains over 72 months. The 1999 measurement value
($\nu_{\mathrm{Rb}}(1999)=6\,834\,682\,610.904\,333\,$Hz) is
conventionally used as reference. A weighted linear fit to the data
(solid line) gives
$\frac{d}{dt}\ln\left(\frac{\nu_{\mathrm{Rb}}}{\nu_{\mathrm{Cs}}}\right)=(-0.5\pm
5.3)\times 10^{-16}\,\mathrm{yr}^{-1}$. MJD represent Modified
Julian Dates.} \label{fig:alphapoint}
\end{figure}
Fig.\ref{fig:alphapoint} summarizes the comparison between $^{87}$Rb
and $^{133}$Cs hyperfine frequencies that have been performed using
the above described fountain ensemble over a duration of 6 years.
Each point on the graph summarizes the result of one to two months
of measurements which include each time an evaluation of all known
systematic effects \cite{Marion2003,Bize1999,Bize01}.
A weighted linear fit to the data in Fig.\ref{fig:alphapoint}
determines how these measurements constrain a possible time
variation of $ \nu_{\mathrm{Rb}}/\nu_{\mathrm{Cs}}$. We find:
\begin{equation} \label{eq:alpha1}
\frac{d}{dt}\ln\left(\frac{\nu_{\mathrm{Rb}}}{\nu_{\mathrm{Cs}}}\right)=(-0.5\pm
5.3)\times 10^{-16}\,\mathrm{yr}^{-1}
\end{equation}
which represents a 100-fold improvement over the Hg$^+$-H hyperfine
energy comparison \cite{Prestage1995}. This results implies the
following constraint:
\begin{equation} \label{eq:alpha2}
\frac{d}{dt}\ln\left(\frac{g_{\mathrm{Cs}}}{g_{\mathrm{Rb}}}\alpha^{0.49}\right)=(0.5\pm5.3)\times
10^{-16}\,\mathrm{yr}^{-1}.
\end{equation}
Using the calculated link between g-factors and $m_{q}$, $m_s$ and
$\Lambda_{\mathrm{QCD}}$ (ref. \cite{Flambaum2003,Flambaum2004}), we
find the following constraint to the variation of fundamental
constants:
\begin{equation} \label{eq:alpha3}
\frac{d}{dt}\ln\left(\alpha^{0.49}\left[m_{q}/\Lambda_{\mathrm{QCD}}
 \right]^{0.174}\left[m_{s}/\Lambda_{\mathrm{QCD}}
 \right]^{0.027}\right)=(0.5\pm5.3)\times 10^{-16}\,\mathrm{yr}^{-1}.
\end{equation}
As pointed out in \cite{Damour1994,Calmet2002,Langacker2002}, the
hypothetical unification of all interactions implies that a
variation of the fine-structure constant $\alpha$ should be
accompanied by a variation of the strong interaction constant and of
elementary particle masses. Within this framework, current estimates
gives \cite{Damour1994,Flambaum2004,Calmet2002,Langacker2002}:
\begin{equation} \label{eq:alpha_unified_theories}
\frac{\delta(m/\Lambda_{\mathrm{QCD}})}{(m/\Lambda_{\mathrm{QCD}})}\sim
35\times\frac{\delta\alpha}{\alpha}.
\end{equation}
Within this theoretical framework, the present comparison between Rb
and Cs fountains (Eq.~\ref{eq:alpha1}) constrains a time variation
of $\alpha$ at the level of $7\times 10^{-17}$~yr$^{-1}$. In the
future, improvement of $^{87}$Rb and $^{133}$Cs fountain to
accuracies of few parts in $10^{16}$ and repeated comparisons over
several years between these devices will improve the above result by
at least one order of magnitude.

The transportable fountain FOM has similarly been used as
a primary standard in the measurement of the frequency
$\nu_{\mathrm{H}}$ of the hydrogen 1S-2S transition performed at
Max-Planck Institut in Garching (Germany)
\cite{Niering2000,Fischer2004}. Two measurements performed over a 4 year
period constrain fractional variations of
$\nu_{\mathrm{Cs}}/\nu_{\mathrm{H}}$ at the level of $(3.2\pm
6.3)\times 10^{-15}$.yr$^{-1}$. This constrains fractional
variations of $g_{\mathrm{Cs}}(m_e/m_p)\alpha^{2.83}$ at the same
level \cite{Prestage1995,Dzuba1999a}.
Combining these results with other recent comparisons
($^{199}$Hg$^+$ optical clock versus $^{133}$Cs fountain
\cite{Bize2003,Udem2001}, $^{171}$Yb$^+$ optical clock versus
$^{133}$Cs fountain \cite{Stenger2001,Peik2004}), it is possible to
independently set limits on variations of $\alpha$,
$g_{\mathrm{Rb}}/g_{\mathrm{Cs}}$ and $g_{\mathrm{Cs}}(m_e/m_p)$.
These measurements test the stability of the electroweak interaction
($\alpha$) and of the strong interaction
($g_{\mathrm{Rb}}/g_{\mathrm{Cs}}$, $g_{\mathrm{Cs}}(m_e/m_p)$)
separately \cite{Fischer2004,Peik2004} and independently of any
cosmological model.

\section{The PHARAO cold atom space clock and ACES}

PHARAO, as ``Projet d'Horloge Atomique par Refroidissement d'Atomes
en Orbite'', has started in 1993 with the objective of performing
fundamental metrology with a space cold atom clock \cite{Laurent1995}.
The combination of laser cooling techniques \cite{lasercooling89}
and microgravity environment indeed allows the development of space
clocks with unprecedented performances.

To demonstrate the feasibility of a compact cold atom clock
operating in microgravity, BNM-SYRTE and LKB with the support of
CNES (the French space agency) have undertaken the construction of a
clock prototype in 1994. The prototype was successfully tested in
1997 in jet plane parabolic flights \cite{Laurent1998}. The same
year, ESA, the European Space Agency, selected the ACES proposal
(Atomic Clock Ensemble in Space) \cite{Salomon1996}. ACES will
perform fundamental physics tests by using the PHARAO cold atom
clock, a H-maser (developed by the Neuch\^{a}tel Observatory) and a time
and frequency transfer system MWL on a platform developed by ESA.
This ensemble will fly on board the international space station in
2008-2009. The station is orbiting at a mean elevation of 400~km
with a 90~mn period and an inclination angle of 51.6$^\mathrm{o}$.
The planned mission duration is 18 months. During the first 6
months, the performances of the PHARAO cold atom clock in space will
be established. Thanks to the microgravity environment the linewidth
of the atomic resonance will be varied by two orders of magnitude
(from 11~Hz to 110~mHz). The target performance is $7\times
10^{-14}\tau^{-1/2}$ for the frequency stability and $10^{-16}$ for
the frequency accuracy. In the second part of the mission, the
onboard clocks will be compared to a number of ground based clocks
operating both in the microwave and the optical domain.

In 2001, PHARAO entered into industrial development under the
management of CNES with the construction of two clock models, an
engineering model for test and validation, and a flight model.

\begin{figure}[htb]
\vspace{80mm}
\caption{The PHARAO sub-systems and interfaces.}
\label{fig:pharao_setup}
\end{figure}

\subsubsection*{The PHARAO instrument}
The clock is composed of four main sub-systems as shown in figure
\ref{fig:pharao_setup}. Each sub-system has been subcontracted to
different manufacturers and they will be assembled at CNES Toulouse
to validate the clock operation.

\begin{figure}[htb]
\vspace{80mm}
\caption{The PHARAO optical bench during assembly. The bench surface
is $55\times 33$~cm$^2$. The laser source includes 8 frequency
stabilized diode lasers. The laser light for atom manipulation is
coupled to the vacuum tube through optical fibers. (courtesy of EADS
SODERN).} \label{fig:pharao_lasersource}
\end{figure}

The laser source provides all the laser tools for cooling, launching
and detection of the atoms. Two extended cavity diode lasers
\cite{Allard04} are used as master lasers. One of them
injection-locks two slave diode lasers to provide high laser power
for capturing $\sim 10^{8}$ in optical molasses. The second laser is
used as a repumping laser. The two master laser frequencies are
stabilized by servo-loops using absorption signals through cesium
cells. The other laser frequencies are synthesized by using 6
acousto-optic modulators (AOM). These AOM also control the laser
beam amplitudes. The output laser beams are connected to the cesium
tube through polarization maintaining optical fibers. Figure
\ref{fig:pharao_lasersource} shows the PHARAO optical bench during
the assembly. The total mass is 20~kg, the volume is 26 liters and
the power consumption is 40~W.

\begin{figure}[htb]
\vspace{80mm}
\caption{Cross-section of the cesium tube. The mass is 45~kg and the
volume 70 liters.} \label{fig:pharao_vacuumtube}
\end{figure}

The cesium tube provides the atomic source, the controlled
environment for the atomic manipulation, the interrogation and
detection process (figure \ref{fig:pharao_lasersource}). Its design
is similar to atomic fountains except for the interrogation zone
where a two zone Ramsey cavity is used. The Ramsey cavity (figure
\ref{fig:pharao_ramseycavity}) has been specially developed for this
application and forms a ring resonator. One coupling system feeds
two symmetrical lateral waveguides which meet at the two interaction
zones. The advantage of this configuration is to provide very weak
phase disturbances of the internal microwave field while enabling
large holes ($8\times 9$~mm) for the atom path. The flight model of
the microwave cavity is currently mounted (September 2004) inside
the atomic fountain FO1 to measure the end to end cavity phase shift
before integration in the flight model. These measurements and
numerical simulations, should enable us to determine the cavity
phase shift effect with an accuracy of a few parts in $10^{17}$.

\begin{figure}[htb]
\vspace{80mm}
\caption{The Ramsey interrogation cavity. The cavity in pure copper
is screwed on a rigid structure  to avoid deformation during launch.
The length of the cavity is $280$~mm. The atoms enter the cavity
through the cut-off waveguide with a rectangular shape (on the
left). Also visible in the center is the the microwave coupling
system. (Courtesy of EADS SODERN and TAS).}
\label{fig:pharao_ramseycavity}
\end{figure}

The cesium tube is designed for a vacuum of $10^{-8}$ Pa in order to
minimize the cold atom losses with the background gas collisions.
Three layers of magnetic shields and a servo system maintain the
magnetic field instability in the interaction zone below 20~pT.
Similarly, the interaction zone temperature is regulated to better
than 0.2$^\mathrm{o}$C.

The microwave chain synthesizes the two radiofrequency signals for
the state selection cavity and the interrogation cavity. A 100~MHz
VCXO (Voltage Control Oscillator) is phase-locked to an Ultra Stable
Oscillator (USO) for the short term stability and to the Space
Hydrogen Maser (SHM) for the medium term. Three USOs have been space
qualified for our application. We have compared these quartz
oscillators to the BNM-SYRTE CSO. Their frequency stability is on
the order of $7\times10^{-14}$ from 1 to 10~s integration time. The
engineering model of the chain has been fully tested and the results
are in agreement with the performance objectives of the space clock.
A further performance verification is currently being made by using
the microwave source in the FO2 fountain. All PHARAO sub-systems are
driven and controlled by a computer (UGB, On Board Data Processing
Unit). The UGB also manages the data flux between the clock and the
ACES payload. When assembled, the clock fills a volume of about
200~l for a weight of 91~kg and an electric consumption of 114~W.

The final assembly of the engineering model of the PHARAO clock
will start at the end of 2005 at CNES-Toulouse. After the clock
functional and performance test are made, the flight model will be
assembled and finally tested. For both models, we expect to reach
$10^{-15}$ frequency accuracy in the Earth gravity environment and
$10^{-16}$ in microgravity environment.

\subsubsection*{Scientific objectives of the ACES mission}

The objectives of PHARAO/ACES are (i) to explore and demonstrate the
high performances of the cold atom space clock (ii) to achieve time
and frequency transfer with stability better than $10^{-16}$ and
(iii) to perform fundamental physics tests. A detailed account can
be found in \cite{Salomon2001}.

The combination of PHARAO with SHM will define an on board frequency
reference having a long term stability and accuracy provided by
PHARAO and a short term stability determined by SHM. The resulting
fluctuations of ACES frequency reference are expected to be about
10~ps per day. The orbit of ISS will allow ground users to compare
and synchronize their own clock to ACES clocks, leading to a
worldwide access to the ultra stable frequency reference of ACES.
The results of these comparisons at $10^{-16}$ level will provide
new tests in fundamental physics such as an improved measurement of
Einstein's gravitational red-shift, a search for a possible
anisotropy of the speed of light and a search for possible
space-time variations of fundamental physical constants, similar to
that described above in section \ref{sec:alpha}. The current most
precise measurement of the red-shift was made by the space mission
Gravitational Probe A (GPA) with an accuracy of $7\times 10^{-5}$
\cite{Vessot1980}. PHARAO/ACES will improve this test by a factor
30. By allowing worldwide comparison between distant clocks,
operating with different atomic species, ACES will play a major role
in establishing new limits for variations of fundamental constants.

Finally, PHARAO/ACES will be a pioneering cold atom experiment in
space. The PHARAO technology can be extended for the development of
a new generation of high performance inertial sensors and clocks
using matter wave interferometry. As for atomic clocks, such sensors
may achieve extremely high sensitivity in micro-gravity environment,
as pointed out in the ESA HYPER project \cite{HYPER2000}. These
instruments could then be used for a large variety of scientific
space missions such as VLBI, gravitational wave detection, and deep
space navigation.

\section{Conclusions}

With methods described in this paper, we expect to bring the
accuracy of $^{133}$Cs fountains at 1 or 2 parts in $10^{16}$. For
$^{87}$Rb, a frequency stability of $1\times 10^{-14}\tau^{-1/2}$
i.e. $3\times 10^{-17}$ at one day seems accessible, together with
an excellent accuracy. Routine operation of these devices over
several years will have a profound impact on ultra-precise time
keeping and fundamental physics tests. To take full benefit of this
performance, long distance time transfer systems must be upgraded.
In particular, the ACES time and frequency transfer system will
enable comparisons at the level of $10^{-16}$ per day in 2008-2009.
Another route currently under study makes use of telecom optical
fibers and over $100$~km  distance a stability of $1\times 10^{-14}$
at 1~s and $2\times 10^{-17}$ at one 1~day has already been
demonstrated \cite{Narbonneau2004}. Extension to larger distances is
under study.

More  generally,  clocks  operating  in  the  optical  domain rather
than  in  the microwave domain are making rapid progress on the
ground \cite{Ekstrom2003}. The frequency of these  clocks  is four
to  five  orders of magnitude higher  than  the  frequency  of
microwave standards and with an equivalent linewidth, the quality
factor of the resonance exceeds that of cesium clocks by the same
factor. Using laser cooled atoms or ions and ultra-stable laser
sources \cite{Young1999}, these optical clocks will likely open the
$10^{-17}-10^{-18}$ stability range. Using the wide frequency comb
generated by femtosecond lasers, it is now possible to connect
virtually all frequency standards together throughout the microwave
to ultra-violet frequency domain \cite{Udem2001,Udem2002}. The
attractive proposal of \cite{Katori2001,Katori2003} to realize an
optical lattice clock is currently receiving a great deal of
interest. In this method, neutral atoms are confined in an optical
lattice in the Lamb-Dicke regime. Light-shifts of the clock levels
induced by the lattice beams are differentially compensated at an
appropriate laser detuning. This proposal combines several
interesting features such as long observation time, large number of
atoms, and recoil-free resonance \cite{Takamoto2003}. Promising
atoms to implement this method are alkaline-earth atoms because of
their strongly forbidden inter-combination line. Ca
\cite{Udem2001,Stenger2004}, Sr \cite{Takamoto2003,Courtillot2003}
and Yb \cite{Kuwamoto1999,Park2003,Porsev2004} are actively studied.

In the frequency stability range of $10^{-17}-10^{-18}$, it is clear
that fluctuations of the Earth potential at the clock location
induced, for instance, by sea tides will affect comparisons  between
distant clocks.  This limitation could  be turned  into  an
advantage if one installs such ultra-stable clock in space where the
gravitational potential can present far reduced fluctuations
compared to ground. As in the past, clocks with very high stability
will have an ever increasing impact on scientific and industrial
applications.

\section{Acknowledgements}
BNM-SYRTE and Laboratoire Kastler Brossel are Unit\'{e}s Associ\'{e}es au
CNRS, UMR 8630 and 8552. This work was supported in part by BNM,
CNRS, CNES and ESA. P. Wolf is on leave from Bureau International
des Poids et Mesures, Pavillon de Breteuil, 92312 S\`{e}vres Cedex,
France. J. Gr\"{u}nert and L. Cacciapuoti acknowledge financial support
from the CAUAC European Research Training Network.


\end{document}